\input{aipcheck}

\documentclass[
    ,final            
  ]
  {aipproc}

\layoutstyle{6x9}

\begin{document}

\title{Heavy-flavor contributions at NNLO in CTEQ PDF analysis}


\author{\footnote{Presented at XIX International Workshop on Deep-Inelastic Scattering and Related Subjects (DIS 2011)
April 11-15, 2011, Newport News VA, USA} Marco Guzzi}{
  address={Department of Physics, Southern Methodist University,
 3215 Daniel Ave. Dallas, TX 75275 USA}
}

\author{Pavel M. Nadolsky}{
  address={Department of Physics, Southern Methodist University,
 3215 Daniel Ave. Dallas, TX 75275 USA}
}

\author{Hung-Liang Lai}{
  address={Taipei Municipal University of Education, Taipei, Taiwan }
}

\author{C.-P. Yuan}{
  address={Dept. of Physics and Astronomy, Michigan State University, E. Lansing, MI 48824, USA}
}

\begin{abstract}
We discuss an NNLO realization of the general mass scheme
S-ACOT-$\chi$ for treatment of heavy-flavour production in 
neutral current deep-inelastic scattering.
Practical implementation of the NNLO calculation is illustrated
on the example of  semi-inclusive structure functions $F_{2 c}(x,Q)$ 
and  $F_{L c}(x,Q)$.
\end{abstract}

\maketitle


\paragraph{\bf Introduction}

Correct computation of heavy-quark contributions to deep-inelastic scattering 
in the global PDF analysis is essential 
for predicting precision cross sections for $W$ and $Z$ 
boson production at the LHC \cite{Tung:2006tb}.

Quark mass effects on DIS cross sections are comparable to next-to-next-to 
leading order (NNLO) contributions, therefore they must 
be included consistently in perturbative computations. 
The S-ACOT-$\chi$ scheme \cite{Collins:1998rz,Kramer:2000hn,Tung:2001mv}
is the default general-mass (GM)
framework of CTEQ global PDF analyses;
it is an improved formulation of the original ACOT scheme 
introduced in \cite{Aivazis:1993kh,Aivazis:1993pi}. 

S-ACOT-$\chi$ is an algorithmic procedure to account for heavy-quark masses 
in DIS structure functions that descends directly from the proof 
of QCD factorization for DIS with massive
quarks \cite{Collins:1998rz}. Compared to the earlier ACOT and S-ACOT schemes, 
the S-ACOT-$\chi$ scheme adds a phenomenologically important requirement 
that all heavy-quark scattering contributions (including those with collinear approximation for heavy quarks) satisfy energy-momentum conservation 
near the quark production threshold.  
In Ref.~\cite{Acotnnlo}, we show that the constraints from energy-momentum conservation 
can be included directly in the QCD factorization theorem. Thus, the S-ACOT-$\chi$ scheme 
is proved to all orders on the same footing as the original ACOT scheme. 

We also present, for the first time, S-ACOT-$\chi$ numerical predictions of NNLO accuracy 
for heavy-quark neutral-current DIS structure functions, $F_{2 c} (x,Q)$  and 
$F_{L c} (x,Q^2)$.

S-ACOT-$\chi$ is attractive for phenomenological applications because of its 
relative simplicity compared to other schemes like 
BMSN \cite{Buza:1996wv} or TR \cite{Thorne:1997ga}.
It achieves a great simplification by replacing 
coefficient functions with incoming heavy-quark lines
by their zero-mass (ZM) expressions, while requiring 
that the light-cone momentum fraction $\xi$ in the PDFs $f_{a,p}(\xi,\mu)$
is always within the interval $\chi<\xi<1$
that satisfies energy-momentum conservation \cite{Tung:2001mv,Nadolsky:2009ge}. 
Here and in the following, a rescaled ``Bjorken-$x$'' variable $\chi=x\, (1+4 m_h^2/Q^2)$
is introduced. This arrangement allows to include all the physical scattering channels, 
while excluding kinematically prohibited $\xi$ values.

The S-ACOT-$\chi$ scheme assumes one value of the number of active quark flavors, $N_f$, 
and one PDF set in each $Q$ bin, in contrast to other possible GM schemes.
Matching of massive 4-flavor predictions on massive 3-flavor predictions at low $Q$, 
and on massless 4-flavor predictions at large $Q$ is realized in S-ACOT-$\chi$ through 
cancellations among certain classes of Feynman diagrams.
These cancellations are improved by using the optimal rescaling variable $\chi$, 
but also hold for other forms of the rescaling variable. 
Practical implementation is easy to handle, and all NNLO Wilson coefficient functions 
that are needed are available from literature.

\paragraph{\bf Summary of the computation}

If the components of inclusive $F_{2,L}(x,Q^2)\equiv F$ are classified 
according to quark couplings to the photon \cite{Forte:2010ta}, $F$ 
can be written as a sum of light and heavy-quark coupled contributions,
\begin{equation}
F=\sum_{l=1}^{N_{l}}F_{l}+F_{h}, 
\end{equation}
where
\begin{eqnarray}
F_{l}=e_{l}^{2}\sum_{a}\left[C_{l,a}\otimes f_{a/p}\right](x,Q),
&&F_{h}=e_{h}^{2}\sum_{a}\left[C_{h,a}\otimes f_{a/p}\right](x,Q)\,.
\label{Fheavy}
\end{eqnarray}
Here $C_{l,h}$ are Wilson coefficient functions, and $f_{a/p}$ are PDFs.
Perturbative expansion up to ${\cal O}(\alpha_s^2)$ leads to
\begin{eqnarray}
F_{h}^{(2)}=e_{h}^{2}\left\{ c_{h,h}^{NS,(2)}\otimes(f_{h/p}+f_{\bar{h}/p})
+C_{h,l}^{(2)}\otimes\Sigma+C_{h,g}^{(2)}\otimes 
f_{g/p}\right\},
\label{Fh}
\end{eqnarray}
where lowercase $c^{(2)}_{a,b}$ and uppercase $C^{(2)_{a,b}}$ represent 
ZM coefficient functions and massive coefficient functions, constructed from results in
Refs.~\cite{vanNeerven:1991nn,Zijlstra:1991qc,Vermaseren:2005qc} 
and Refs.~\cite{Buza:1996wv,Buza:1995ie,Riemersma:1994hv,Laenen:1992zk}, respectively. 
$\Sigma$ is the singlet PDF combination. 
A detailed derivation of these coefficient functions is presented in Ref.~\cite{Acotnnlo}.
A similar expression for $F_{l}$ is given by 
\begin{eqnarray}
F_{l}^{(2)} = e_{l}^{2}\left\{C_{l,l}^{NS,(2)}\otimes(f_{l/p} + f_{\bar{l}/p})
+c^{PS,(2)}\otimes\Sigma+c_{l,g}^{(2)}\otimes f_{g/p}\right\},
\label{Fl}
\end{eqnarray}
where $c_{l,g}^{(2)}$, $c^{PS,(2)}$ and $C_{l,l}^{NS,(2)}$ are constructed from components 
available in Refs.~\cite{vanNeerven:1991nn,Zijlstra:1991qc,Vermaseren:2005qc,Acotnnlo}.

\vspace{-0.3cm}

\begin{figure}
\includegraphics[height=.35\textheight]{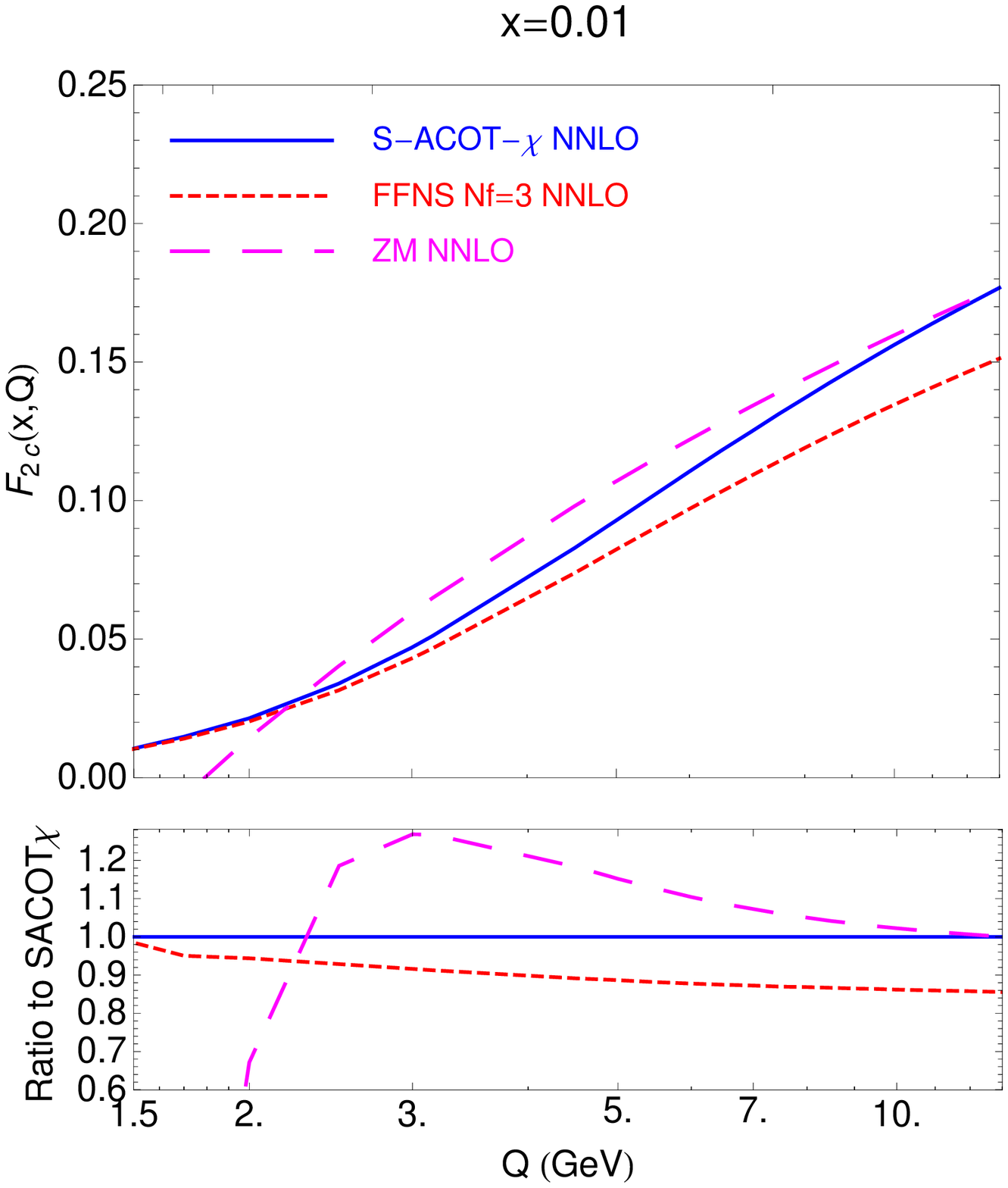}\hspace{-3cm}
\includegraphics[height=.35\textheight]{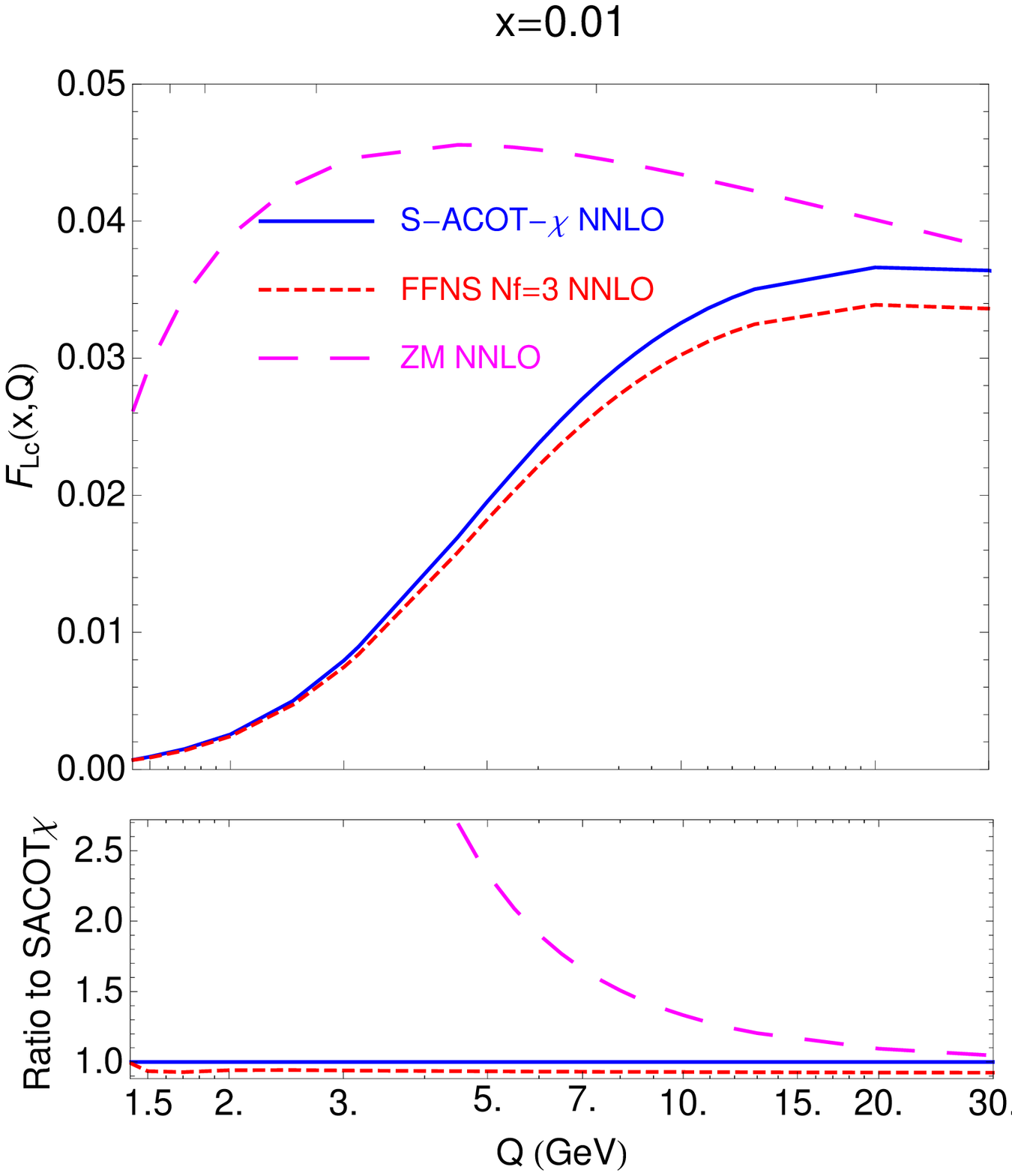}
\caption{Upper insets: semi-inclusive $F_{2c}(x,Q)$ and $F_{Lc}(x,Q)$ at NNLO
as a function of $Q$ at $x=10^{-2}$. 
Lower insets: ratios of ZM and FFN predictions to the respective S-ACOT-$\chi$ predictions.}
\label{Q_dep}
\end{figure}

\paragraph{\bf Numerical results}
In the following, we show representative NNLO predictions for $F_{2c}$ and $F_{Lc}$,
computed using Les Houches toy PDFs \cite{Giele:2002hx,Whalley:2005nh} that are evolved
with 4 active flavors by HOPPET computer code \cite{Salam:2008qg}. 
In Fig.\ref{Q_dep}, NNLO 
S-ACOT-$\chi$ predictions for   $F_{2c}$ (left panel) and $F_{Lc}$ (right panel) 
are shown vs. $Q$ by blue solid lines. ZM 4-flavor 
predictions at NNLO are shown by purple long-dashed lines, 
and 3-flavor (fixed-flavor-number, or FFN) predictions by red short-dashed lines.
Fig.\ref{Q_dep} shows that S-ACOT-$\chi$ smoothly interpolates in between
the FFN and ZM schemes, and that it 
reduces to the FFN scheme at $Q \approx m_c$ and to the ZM scheme when $Q \gg m_c$.
Lower insets show ratios of the FFN and ZM predictions to the respective S-ACOT-$\chi$
predictions, to elucidate differences in the intermediate region. 

\begin{figure}
\includegraphics[height=.30\textheight]{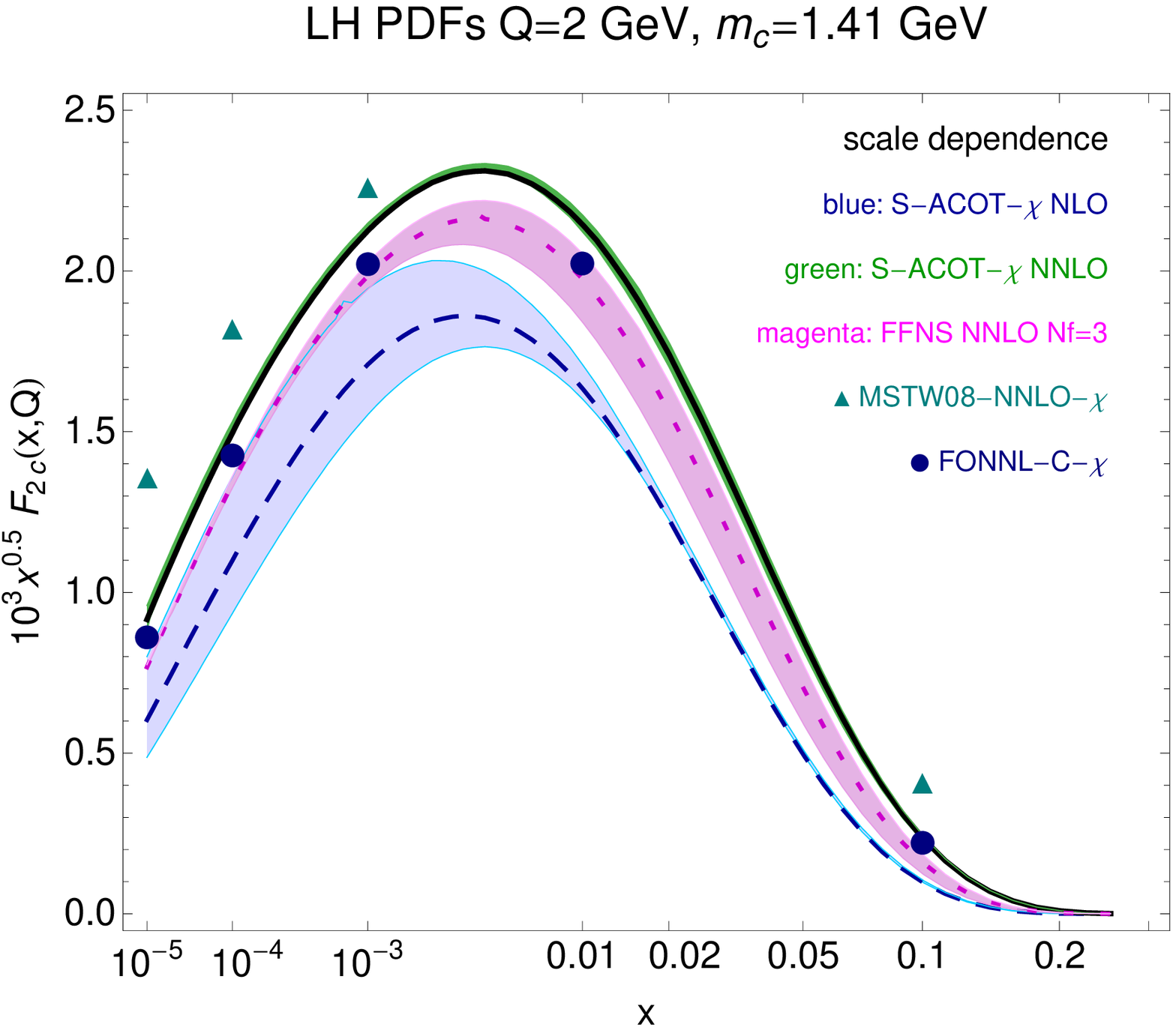}
\includegraphics[height=.30\textheight]{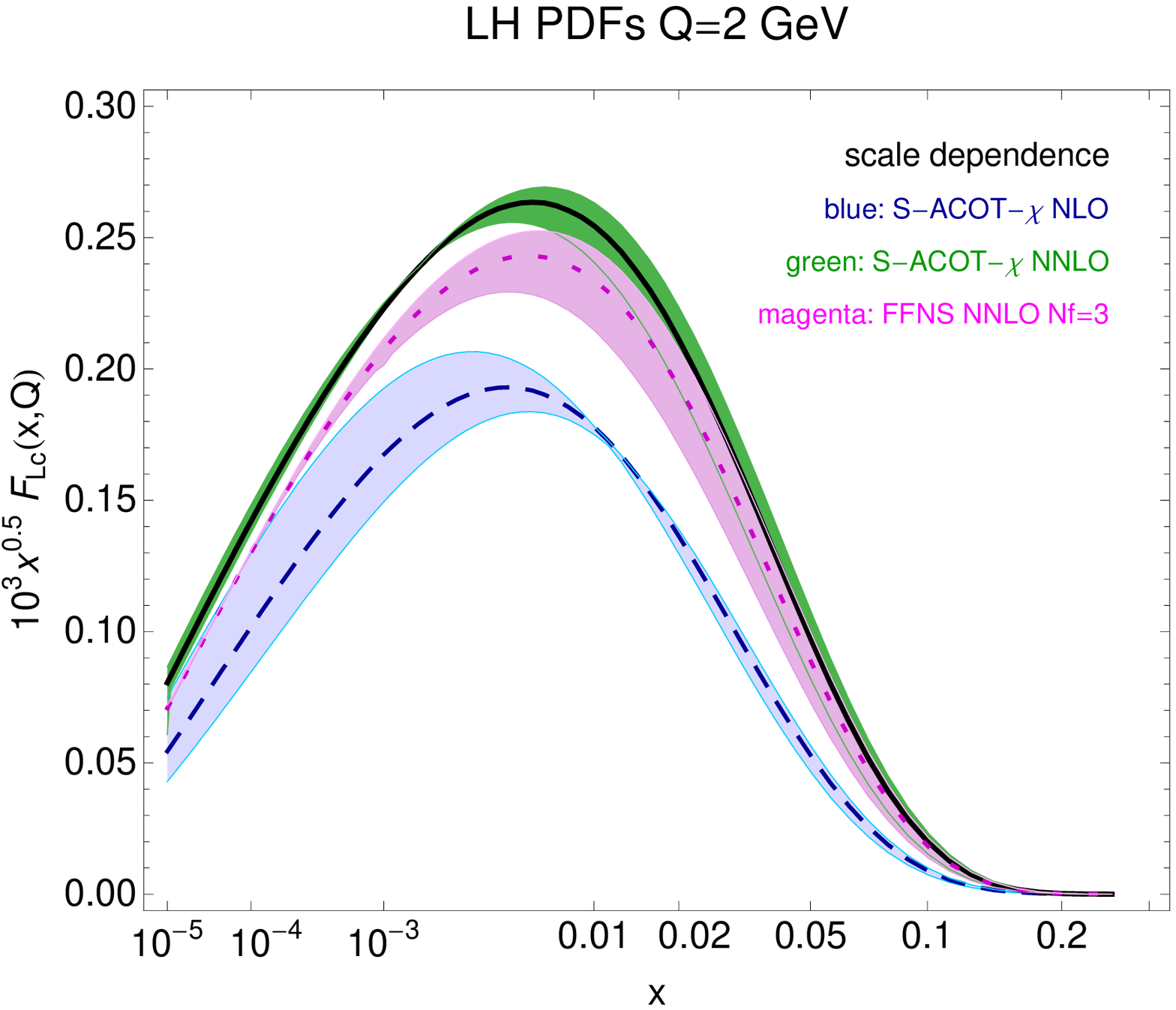}
\caption{Factorization scale dependence for semi-inclusive $F_{2,L c}(x,Q)$ 
at NNLO as a function of $x$ at $Q=2$ GeV.}
\label{scl_dep}
\end{figure}

Fig.~\ref{scl_dep} shows a remarkable reduction of the factorization 
scale dependence of $F_{2,L c}$. 
Without extra tuning of the factorization scale, 
the S-ACOT-$\chi$ prediction is close to FFN and other NNLO schemes at $Q\approx m_c$.
Here scattered symbols in the left panel correspond to predictions based on 
MSTW/TR' NNLO coefficient functions (sea-green triangles) \cite{Martin:2009iq} and 
FONLL-C NNLO coefficient functions (blue circles) \cite{Forte:2010ta}.
The solid black line is 
the S-ACOT-$\chi$ NNLO prediction corresponding to a reference factorization 
scale $\mu=\sqrt{Q^2 + m_c^2}$.
The green band is the theoretical uncertainty in the S-ACOT-$\chi$ prediction 
due to the variation of the scale in the range $Q< \mu <\sqrt{Q^2 + 4 m_c^2}$.
The purple band represents scale variations in the FFN prediction at NNLO
around the reference scale values indicated by the magenta short-dashed line.
The light blue band around the blue-dashed line 
represents the S-ACOT-$\chi$ prediction at NLO and its scale uncertainty.  

\vspace{-0.3cm}

\paragraph{\bf Conclusions}
We have shown that the S-ACOT-$\chi$ scheme is fully compatible with the QCD factorization theorem, and that an NNLO computation of $F_{2,L c}$
in the S-ACOT-$\chi$ scheme is viable.
S-ACOT-$\chi$ predictions at NNLO are stable and show significant
reduction in  the factorization scale dependence 
(see Fig.\ref{scl_dep}), compared to NLO computations. 
This is the most challenging component of the CTEQ global analyses at NNLO.
A full description of NNLO S-ACOT-$\chi$ computations is available in \cite{Acotnnlo}.

\vspace{-0.5cm}

\begin{theacknowledgments}
This work was supported in part by the U.S. Department of Energy under Grant DE-AC02-
06CH11357 and by the U.S. National Science Foundation under Grant No. PHY-0855561, 
and by the National Science Council of Taiwan under Grant NSC-98-2112-M-133-002-MY3.
\end{theacknowledgments}
\vspace{-0.5cm}


\bibliographystyle{aipproc}   


\hyphenation{Post-Script Sprin-ger}

\hyphenation{Post-Script Sprin-ger}

\IfFileExists{\jobname.bbl}{}
 {\typeout{}
  \typeout{******************************************}
  \typeout{** Please run "bibtex \jobname" to optain}
  \typeout{** the bibliography and then re-run LaTeX}
  \typeout{** twice to fix the references!}
  \typeout{******************************************}
  \typeout{}
 }

\end{document}